%% file: 0-main.tex
\documentclass[sigconf]{acmart}

\renewcommand\footnotetextcopyrightpermission[1]{} 
\makeatletter
\def\@ACM@checkaffil{
    \if@ACM@instpresent\else
    \ClassWarningNoLine{\@classname}{No institution present for an affiliation}%
    \fi
    \if@ACM@citypresent\else
    \ClassWarningNoLine{\@classname}{No city present for an affiliation}%
    \fi
    \if@ACM@countrypresent\else
        \ClassWarningNoLine{\@classname}{No country present for an affiliation}%
    \fi
}
\makeatother
\AtBeginDocument{%
  \providecommand\BibTeX{{%
    \normalfont B\kern-0.5em{\scshape i\kern-0.25em b}\kern-0.8em\TeX}}}

\setcopyright{acmcopyright}

\copyrightyear{2024} 
\acmYear{2024} 
\setcopyright{acmlicensed}\acmConference[XXX '24]{}{June 4-7, 2024}{XXX}
\acmBooktitle{The 15th ACM International Conference on Future Energy Systems (e-Energy '24), June 4-7, 2024,Singapore}
\acmPrice{15.00}
\acmDOI{XXXXXXX}
\acmISBN{XXXX}

\usepackage{float}
\usepackage[utf8]{inputenc}
\usepackage{url}
\usepackage{multirow}

\usepackage{amssymb}
\usepackage{natbib} 
\usepackage{color}
\usepackage{xcolor}
\usepackage{appendix}
\usepackage{verbatim}
\usepackage{algorithm, algorithmic}
\usepackage{amsmath}
\usepackage{xspace}
\usepackage{subfigure}
\usepackage{diagbox}

\newcommand{\ignore}[1]{{}}

\newtheorem{theorem}{Theorem}

\usepackage{listings}
\definecolor{keyword-green}{rgb}{0.2, 0.5, 0.3}
\definecolor{string-red}{rgb}{0.7, 0.1, 0.2}
\definecolor{important-red}{rgb}{0.8, 0.25, 0.2}
\definecolor{comment-grey}{rgb}{0.4, 0.5, 0.6}
\definecolor{item-blue}{rgb}{0.25, 0.3, 0.85}
\definecolor{text-blue}{rgb}{0.25, 0.2, 0.4}
\definecolor{number-black}{rgb}{0.1, 0.1, 0.15}
\definecolor{another-orange}{rgb}{0.7, 0.5, 0.15}

\lstdefinestyle{sql-style}{
language=SQL,
commentstyle=\color{comment-grey},
keywordstyle=\color{keyword-green}\bfseries,
stringstyle=\color{string-red},
numbers=left,
numbersep=5pt,
numberstyle=\color{number-black}\tiny,
frame=lines,
breaklines=true,
basewidth={.5em},
morekeywords={FILTER},
moredelim=**[is][\color{item-blue}]{~}{~},
framextopmargin=0.5em,
framexbottommargin=0.5em,
}

\lstdefinestyle{py-style}{
language=Python,
commentstyle=\color{comment-grey},
keywordstyle=\color{keyword-green}\bfseries,
stringstyle=\color{string-red},
numbers=left,
numbersep=5pt,
numberstyle=\color{number-black}\tiny,
frame=lines,
breaklines=true,
basewidth={.5em},
morekeywords={FILTER},
moredelim=**[is][\color{item-blue}]{~}{~},
moredelim=**[is][\color{item-blue}\bfseries]{~~}{~~},
moredelim=[is][\color{string-red}]{--}{--},
framextopmargin=0.5em,
framexbottommargin=0.5em,
}

\lstdefinestyle{sparql-style}{
language=sparql,
commentstyle=\color{commend-grey},
keywordstyle=\color{keyword-green}\bfseries,
stringstyle=\color{string-red},
numbers=left,
numbersep=5pt,
numberstyle=\color{number-black}\tiny,
rulecolor=\color{number-black},
frame=lines,
breaklines=true,
basewidth={.5em},
morekeywords={FILTER},
moredelim=**[is][\color{item-blue}\bfseries]{~}{~},
moredelim=**[is][\color{keyword-green}\bfseries]{!}{!},
moredelim=**[is][\color{important-red}\bfseries]{=}{=},
moredelim=**[is][\color{important-red}]{==}{==},
moredelim=**[is][\bfseries]{!!}{!!},
framextopmargin=0.5em,
framexbottommargin=0.5em,
}

\usepackage{acmart-taps}

\begin{document}

\title{Embodied Carbon Accounting through Spatial-Temporal Embodied Carbon Models}

\author{Xiaoyang Zhang}
\affiliation{%
  \institution{The Hong Kong Polytechnic Univ.}
}
\email{xiaoyang.zhang@connect.polyu.hk}

\author{Yijie Yang}
\affiliation{%
  \institution{The Hong Kong Polytechnic Univ.}
}
\email{yi-jie.yang@connect.polyu.hk}

\author{Dan Wang}
\affiliation{%
  \institution{The Hong Kong Polytechnic Univ.}
}
\email{dan.wang@polyu.edu.hk}

\renewcommand{\shortauthors}{Zhang et al.}

\begin{abstract}

\textit{Embodied carbon} is the total carbon released from the processes associated with a product from cradle to gate. In many industry sectors, embodied carbon dominates the overall carbon footprint. \textit{Embodied carbon accounting}, i.e., to estimate the embodied carbon of a product, has become an important research topic. 

Existing studies derive the embodied carbon through life cycle analysis (LCA) reports. Current LCA reports only provide the carbon emission of a \textit{product class}, e.g., 28nm CPU, yet a \textit{product instance} can be manufactured from diverse regions and in diverse time periods, e.g., a winter period of Ireland (Intel). It is known that the carbon emission depends on the electricity generation process which has spatial and temporal dynamics. Therefore, the embodied carbon of a specific product instance can largely differ from its product class. In this paper, we present new spatial-temporal embodied carbon models for embodied carbon accounting. We observe significant differences between current embodied carbon models and our spatial-temporal embodied carbon models, e.g., for 7nm CPU the difference can be 13.69\%.

\end{abstract}

\maketitle
\input{1-Introduction}
\input{3-Problem}

\input{5-Evaluation}
\input{6-Conclusion}

\bibliographystyle{ACM-Reference-Format}
\bibliography{ref}


\input{7-Appendix}


\end{document}

%% file: 1-Introduction.tex
\section{Introduction}
In recent years, there is growing awareness on sustainability \cite{hottenroth2022beyond,ragwitz2023scenarios} and carbon reduction \cite{lin2023adapting,sukprasert2023spatiotemporal,lin2021evaluating,lin2023reducing}. \textit{Embodied carbon} is the total carbon released from the processes associated with a product from cradle to gate \cite{hammond2011embodied}. In many industry sectors, embodied carbon dominates the overall carbon footprint of a product as compared to its \textit{operational carbon} \cite{bashir2023promise}. For example, the embodied carbon of iPhone 11 accounts for 79\% of its overall carbon footprint \cite{gupta2022act}.

\textit{Embodied carbon accounting}, i.e., to estimate the embodied carbon of a product, has become an important research topic \cite{chien2023reducing,ulrich2022comparison,chien2023genai,maji2023untangling}. There are studies on embodied carbon accounting for computer hardware of CPU, memory, and storage \cite{tannu2022dirty,gupta2022act}. The methodology is to leverage life cycle analysis (LCA) reports \cite{menoufi2011life}. For example, in the Environmental, Social, and Governance (ESG) report of SK hynix, the embodied carbon of memory (LPDDR4) is 48 g/GB \cite{SK}. To increase the embodied carbon accounting accuracy, current studies (1) derive the embodied carbon of a product from the carbon of its manufacturing processes. For example, ACT \cite{gupta2022act} carefully models the carbon emissions of the manufacturing process of CPU; and/or (2) analyze the LCA reports from diverse sources. For example, the embodied carbon of storage was estimated \cite{tannu2022dirty} from LCA reports of both a computer vendor and a storage manufacturer.

The embodied carbon of a product heavily depends on the \textit{carbon intensity} \cite{unnewehr2022open,maji2022dacf,zhang2023gnn} of the electricity used in the manufacturing process of this product. Specifically, carbon intensity is the amount of carbon emitted when generating a unit of electricity; and different energy sources, e.g., coal or solar, can lead to different carbon emissions when generating a unit of electricity. The carbon intensity of electricity has spatial and temporal dynamics. The spatial dynamics come from the energy policies of the geographic locations. For example, the electricity generated in Taiwan has higher carbon intensity as compared to Ireland, since the energy policy of Taiwan relies on traditional energy sources due to its lack of renewable energy sources. The temporal dynamics come from the environmental dynamics, which affect the amount of renewable energy sources when generating electricity \cite{schmeck2022energy,meisenbacher2023autopv}. For example, the electricity generated in the winter of Ireland has less carbon intensity as compared to summer, since the wind sources are abundant in the winter of Ireland.

None of the existing studies on embodied carbon accounting has taken the spatial and temporal dynamics into consideration. Existing LCA reports on the embodied carbon of a product represent a \textit{product class}, e.g., 28nm CPU, with the same manufacturing process. Yet a \textit{product instance} can be manufactured from diverse regions and in diverse time periods, e.g., in the winter of Ireland (Intel) or in the summer of Taiwan (TSMC).

In this paper, we present new embodied carbon models that can extend existing embodied carbon models to capture the spatial and temporal dynamics in different granularity. Specifically, we observe that (1) the energy policies have granularity in a country-level, in a treaty-zone-level of multiple countries with energy treaties, e.g., European Union (EU), Association of Southeast Asian Nations (ASEAN), etc.; and in a global-level and (2) the environmental dynamics have granularity in a day-level, in a season-level; and in a year-level. Embodied carbon models at different granularity will be useful for different applications. For example, manufacturers with regular purchasing of computer hardware for their assembly lines may be concerned with embodied carbon on a year-level. Consumers with relatively flexible needs may be more concerned about the embodied carbon with finer granularity on season-level or even day-level, e.g., individual consumers with environmental protection awareness \cite{pandey2011carbon}. Some carbon-related trading parties may be concerned with the embodied carbon at the treaty-zone-level. For example, the EU will impose a carbon tax on non-EU products according to the Carbon Border Adjustment Mechanism (CBAM).

The existing models can be considered as a spatial-temporal embodied carbon (STEC) model of a global-level and a year-level (STEC-GY). In this paper, we additionally study (1) an STEC model of a country-level and a day-level (STEC-CD), (2) an STEC model of a country-level and a season-level (STEC-CS), and (3) an STEC model of a zone-treaty-level and a year-level (STEC-ZY) for computer hardware of CPU, memory, and storage. We study these models due to the availability of data. Note that one challenge to develop our models is to collect public data, which are spread in various reports. We made efforts to collect and \textit{organize} data. We plan to open source the data when this paper goes public.

We compare our models with ACT, a state-of-the-art embodied carbon model on CPU, memory, and storage. We observe significant differences in the embodied carbon accounting with and without taking the spatial-temporal factors into consideration. 
Firstly, the differences exist and are significant at any granularity (e.g., 13.69\% for 7nm CPU at the country-level and season-level). Second, we find there are more dynamics on finer granularity. For example, the average maximum difference of STEC-CD is greater than STEC-ZY (33.62\% compared to 19.29\%). Last, we observe that embodied carbon in some countries has seasonal patterns and can be significantly affected by extreme weather.



%% file: 3-Problem.tex
\section{Spatial-Temporal Embodied Carbon (STEC) Models}
\subsection{A Framework for STEC Models}


We can classify embodied carbon models according to different spatial and temporal granularity (see Table \ref{model}). Along the spatial dimension, there is granularity on country-level, treaty-zone-level, and global-level. Along the temporal dimension, there is granularity on day-level, season-level, and year-level. Different models can serve different applications. For instance, STEC-CD is beneficial for consumers who have the flexibility to adjust their behaviors and purchase computer hardware. STEC-GY is suitable for consumers who make regular purchases to feed their assembly lines. 

The key difficulty is to collect data on renewable energy sources at different granularity. We made an effort in data collection so that we can study STEC-CD/CS/ZY; and we omit STEC-CY due to space limitations. Current data cannot support us to study STEC-TD/TS/GD/GS. Note that although we can study STEC-CD, it does not mean that we can aggregate the data to develop STEC-TD. This is because we only have the day-level data of two countries, Italy and Ireland, thus we cannot study STEC-TD, e.g., at an EU-level.

\begin{table}[htbp]
\caption{Classification of embodied carbon models based on different spatial and temporal granularity}
\label{model}
\scalebox{0.9}{
\begin{tabular}{c|c|c|c}
\toprule
\diagbox{Spatial}{Temporal} & Day-level & Season-level& Year-level \\
\hline 
Country-level & STEC-CD &STEC-CS&-\\
\hline 
Treaty-zone-level & $\times$ & $\times$ &STEC-ZY\\
\hline
Global-level & $\times$ & $\times$ &\cite{gupta2022act}\cite{kohler2023carbon}\cite{tannu2022dirty}\\
\toprule
\end{tabular}
}
\vspace{-1.5em}
\end{table}

\subsection{The Architectural CO$_2$ Tool (ACT) Model \cite{gupta2022act}}

We first present a state-of-the-art embodied carbon model, the Architectural CO$_2$ Tool (ACT) Model \cite{gupta2022act}. ACT can be considered as an STEC-GY model. It is developed for three fundamental computer hardware, CPU, memory (e.g., DRAM), and storage (e.g., SSD, HDD), denoted as $EC_C, EC_M, EC_S$. Different hardware models (e.g., 7nm CPU vs. 28nm CPU) have different embodied carbon. The embodied carbon is estimated per-unit size ($CPS$) of the respective computer hardware; specifically, per-unit die size (in cm$^2$) for CPU and per-unit storage size (in Gigabytes/GB) for memory and storage.


\textbf{Memory and Storage:} In the memory and storage industry, companies have developed a convention to release their embodied carbon, i.e., $CPS$, in their annual ESG reports. Note that different manufacturing processes lead to different $CPS$. For example, a 6GB  DRAM memory using the manufacturing process of LPDDR4 from SK hynix has $CPS = 47.5 g/GB$ (released by the SK hynix report 2021 ), and a 1TB SSD storage of the series of Nytro 1551 from Seagate has $CPS = 3.95 g/GB$ (released by the Seagate product report 2019). Thus, $EC_M = 47.5 g/GB$ and $EC_S = 3.95 g/GB$ for these types of memory and storage, respectively.

\textbf{CPU:} In the microprocessor industry, companies do not directly release their embodied carbon. As such, ACT estimates the embodied carbon by modeling the carbon emission during the manufacturing of microprocessors. Such a manufacturing process has three \textit{carbon emission components}: (1) the carbon released by the gas in the manufacturing process, specifically, from the fluoridated compounds (e.g., SF6, NF3, CF4, CHF3), denoted as $GPS$ (gas per-size), (2) the carbon released by procuring raw materials, denoted as $MPS$ (material per-size); and (3) the carbon released by the electricity consumed during the manufacturing process of the microprocessor. This is calculated by electricity consumption per unit of size, denoted as $EPS$,  and \textit{carbon intensity} of a power grid, denoted by $CI$. $CI$ is the amount of carbon emitted when generating a unit of electricity and it can be obtained by the reports of a power grid \cite{zhang2023gnn}. Both GPS and EPS can be obtained from the related research paper \cite{bardon2020dtco}, and MPS can be obtained from industrial research reports \cite{boyd2011life}.

The embodied carbon $EC_C$ is computed according to Eq. \ref{eq: ce_cpu_act}.

\begin{center}
\begin{equation}
EC_C = (GPS + MPS + CI \times EPS) \times \frac{1}{Y}
\label{eq: ce_cpu_act}
\end{equation}
\end{center}
Here $Y$ is the fabrication yields, e.g., 95\% for TSMC 14nm CPU.

\subsection{Spatial-Temporal Carbon Intensity}

The key to developing an STEC model is to study the three carbon emission components in the manufacturing process in spatial and temporal dimensions. For GPS and MPS, we conjecture that they are less dynamic in spatial and temporal dimensions since they are less related to renewable energy, though we admit that further investigation should be carried out. 

The carbon emissions from the electricity component have a greater contribution. For example, for a 7nm CPU, the carbon emission contributions are GPS 11\%, MPS 28\%, and EPS 61\%. 
Intrinsically, carbon emissions in electricity depend on the energy sources used to generate a unit of electricity, e.g., solar or coal; as well as the amount of renewable energy, e.g., solar. The former has diversity according to locations and the latter has diversity according to weather and seasons. There are studies on the spatial and temporal carbon emissions in electricity and we adopt the model in \cite{maji2022carboncast}.

Here, the carbon intensity $CI(s,t)$ is calculated according to a specific time $t$ and location $s$. Let $\mathcal{E}$ be the set of energy sources. Let $E^k(s,t)$ be the electricity generated by source $k$ at time $t$ and location $s$. Clearly $E(s,t) = \sum_{k \in \mathcal{E}} E^k(s,t)$. The carbon emitted by each type of energy source in electricity generation differs. Let $ef^k$ be the \textit{carbon emission factor} of an energy source $k$. We show the carbon emission factors in Table \ref{table: ef} in the appendix. Finally, the carbon intensity of electricity is the ratio of the total carbon emissions as against the total electricity generated ($E(s,t)$).  

\begin{center}
\begin{equation}
CI(s,t) = \frac{\sum_{k\in{\mathcal{E}}} ef^k \times E^k(s,t)}{E(s,t)}
\label{eq: ci}
\end{equation}
\end{center}

\subsection{Spatial-Temporal Embodied Carbon Models}
We now present our models. We capture the spatial and temporal dynamics in the carbon intensity $CI(s,t)$, and the embodied carbon can be calculated by $CI(s,t)$ with the electricity consumption. Recall that the memory and storage models $EC_M$, $EC_S$ in \cite{gupta2022act} directly apply the embodied carbon from corporate annual reports. To develop our spatial and temporal models, we need the electricity consumption; which are difficult to find in these annual reports. Therefore, we first reversely compute the electricity consumption out of their embodied carbon.

We first develop the basic models on memory, storage, and CPU ($EC_M(s,t), EC_S(s,t), EC_C(s,t)$). We then develop STEC-CD, STEC-CS, and STEC-ZY, where we differentiate different granularity on $s$, e.g., country-level, treaty-zone-level of multiple countries, and $t$, e.g., day, season.

\subsubsection{Basic models.} 
The three models are illustrated as follows:

\textbf{Memory:}
 The spatial-temporal embodied carbon of the memory ($EC_m(s,t)$) contains two components: (1) the carbon released by the electricity consumed during the manufacturing process of the memory. It is calculated by multiplying the electricity consumption per unit of size ($EPS$) by the carbon intensity of a power grid ($CI(s,t)$); and (2) the carbon released independent of electricity, such as raw materials, distribution, and packaging, donated as $\alpha_M$. The calculation of $\alpha_M$ involves yearly $EC_M$, $EPS$, $BD$ (bit density), along with the annual carbon intensity $CI$, all of which can be obtained from the reports. This relationship is depicted in Eq. \ref{eq:alpha_M}.

\begin{equation}
\alpha_M = EC_M - CI \times EPS \div BD
\label{eq:alpha_M}
\end{equation}
As such, $EC_M(s,t)$ is calculated by Eq. \ref{eq: EC_DRAM}.

\begin{equation}
\begin{aligned} 
EC_{M}(s,t) = CI(s,t) \times EPS  \div BD  + \alpha_M
\label{eq: EC_DRAM}
\end{aligned} 
\end{equation}

\textbf{Storage:} 
The spatial-temporal embodied carbon of the storage ($EC_S(s,t)$) contains two components: (1) the carbon emissions from the manufacturing process of storage, which is determined by multiplying the electricity consumed during manufacturing ($EPG$) by the annual carbon intensity $CI$; and (2) the carbon released independent of electricity such as raw materials, distribution, and packaging, donated as $\alpha_S$. $\alpha_S$ can be found in the industry reports. $EPG$ can be calculated through annual $EC_S$ in the reports and annual carbon intensity $CI$, which is shown in Eq. \ref{eq:epg}. 
\begin{equation}
EPG = (EC_S-\alpha_S) / CI
\label{eq:epg}
\end{equation}

As such,  $EC_S(s,t)$ can be calculated by Eq. \ref{eq: EC_storage}.
\begin{equation}
EC_{S}(s,t) = CI(s,t) \times EPG + \alpha_S
\label{eq: EC_storage}
\end{equation}

\textbf{CPU:} The spatial-temporal embodied carbon of CPU ($EM_C(s,t)$) can be calculated using $CI(s,t)$ instead of $CI$ in Eq.\ref{eq: ce_cpu_act}, as Eq. \ref{eq: ce_cpu} shows.

\begin{equation}
\label{eq: ce_cpu}
EC_C(s,t) = GPS + MPS + CI(s,t) \times EPS
\end{equation}

\subsubsection{STEC-CD, STEC-CS, and STEC-ZY}

In the STEC-CD model, $t \in \{ day_i , \forall i \in [1, 365]$ \}, $s \in \{country\}$. $ \{country\}$ is the set of all the CPU production places, e.g., Taiwan, Korea, USA, etc.

In the  STEC-CS model, $t \in \{ spring, summer, fall, winter\}$, $s \in \{country\}$. $ \{country\}$ is the set of all the CPU production places.

In the  STEC-ZY model, $t \in \{ year\}$,  $s \in \{zone\}$. $\{ year\}$ is the set of the production years. $ \{zone\}$ is the set of treaty-zone-level of multiple countries with energy treaties, e.g., ASEAN, EU, etc.

%% file: 5-Evaluation.tex
\section{Evaluation}

In this section, we compare STEC models with STEC-GY and we compare on three types of hardware, CPU, Memory, and Storage. 

Table \ref{table: datasource} (a) shows the data sources for electricity. One-World-in-Data has year-level data for two zones (EU and ASEAN) and countries. EMBIR has month-level data for countries. ENTSOE has day-level data for two countries. Table \ref{table: datasource} (b) shows the data sources for the two other components of carbon emissions. 


\subsection{The Evaluation of STEC-CD}

\begin{figure}[]
    \centering
    \includegraphics[width=0.46\textwidth]{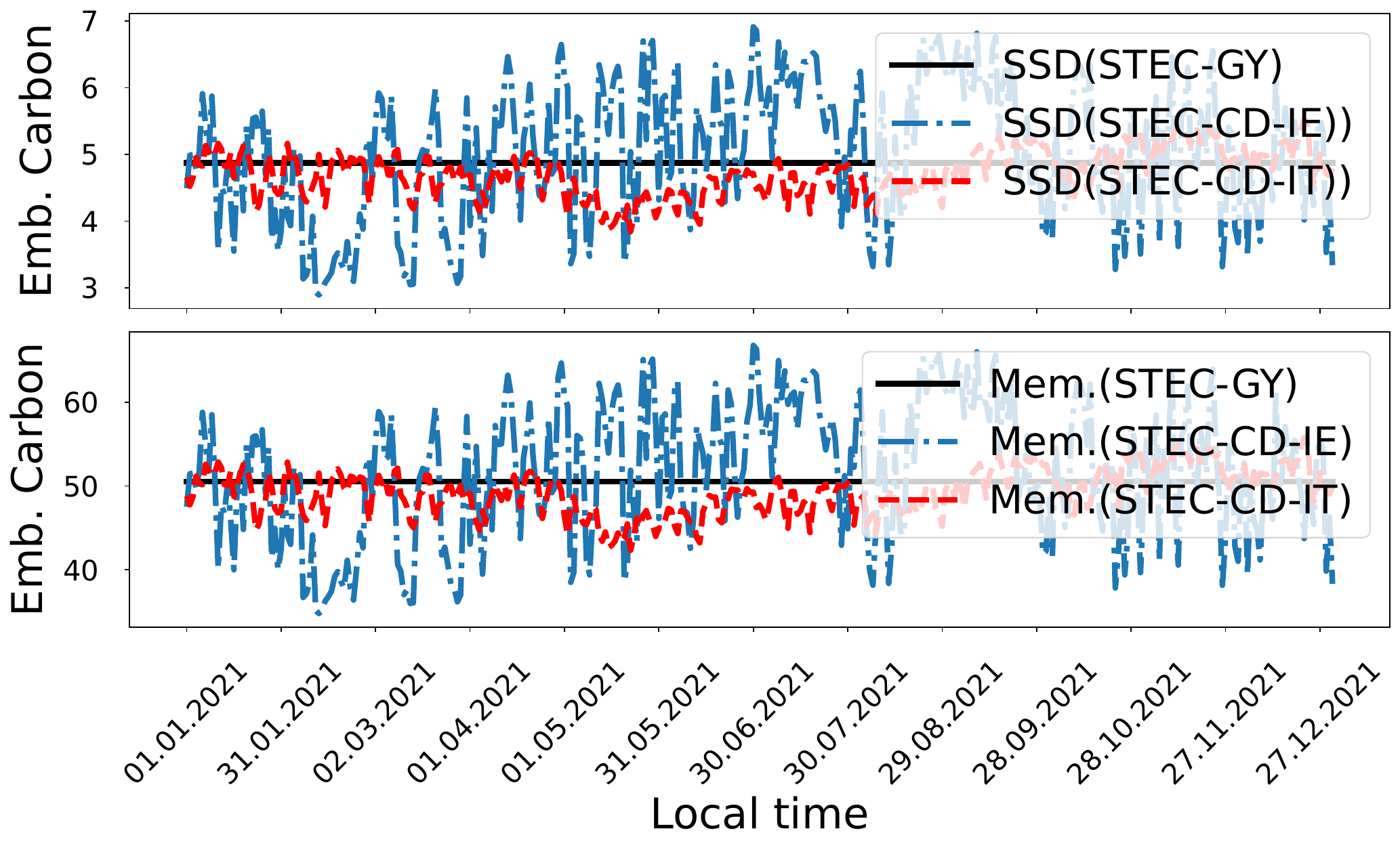}
    \vspace{-1.5em}
    \caption{The comparison between STEC-CD and STEC-GY.}
    \label{fig: STEC-CD}
    \vspace{-1em}
\end{figure}

Table \ref{table: stec-cd} presents the comparison between STEC-CD and STEC-GY (baseline) in the countries with fine-grained data available (Ireland and Italy). We can find the following. (1) The difference in the performance between STEC-CD and STEC-GY is significant. Specifically, the average difference is 9.44\%, and the average maximum difference is 33.62\%. (2) The SSD has a greater difference than HDD (11.76\% compared to 7.15\%). The reason is that SSD is manufactured by a more advanced process than HDD, which results in more electricity consumption and then brings more dynamics. Therefore, advanced processes will amplify the difference in embodied carbon. We conduct an extended evaluation of the embodied carbon on the CPUs from 3nm to 28nm to show this trend as Fig. \ref{fig: spitical} shows in the appendix. (3) The proportion of variable renewable energy sources (VRE) in electricity can affect the dynamics of the embodied carbon. For example, there are more dynamics on the embodied carbon of the product manufactured in Ireland than in Italy, as Fig. \ref{fig: STEC-CD} shows. Intrinsically, there is more VRE in the grids of Ireland than in Italy, which brings more dynamics for the embodied carbon.


\begin{table}[]
\caption{The comparison between STEC-CD and STEC-GY on CPU (7nm), memory (10nm DDR4), SSD \cite{ssd}, and HDD \cite{hdd}.}
\label{table: stec-cd}
\vspace{-1.5em}
\begin{tabular}{ccc}
\toprule
Hardware & Ave. Difference (\%) & Max. Difference (\%) \\ \toprule
CPU & 9.80 & 34.92 \\
SSD & 11.76 & 41.9 \\
HDD & 7.15 & 25.48 \\
Memory & 9.03 & 32.17 \\ \hline
Average & 9.44 & 33.62 \\ \toprule
\end{tabular}
\vspace{-1.5em}
\end{table}

\subsection{The Evaluation of STEC-CS}

\begin{figure}[htbp]
    \centering
    \includegraphics[width=0.46\textwidth]{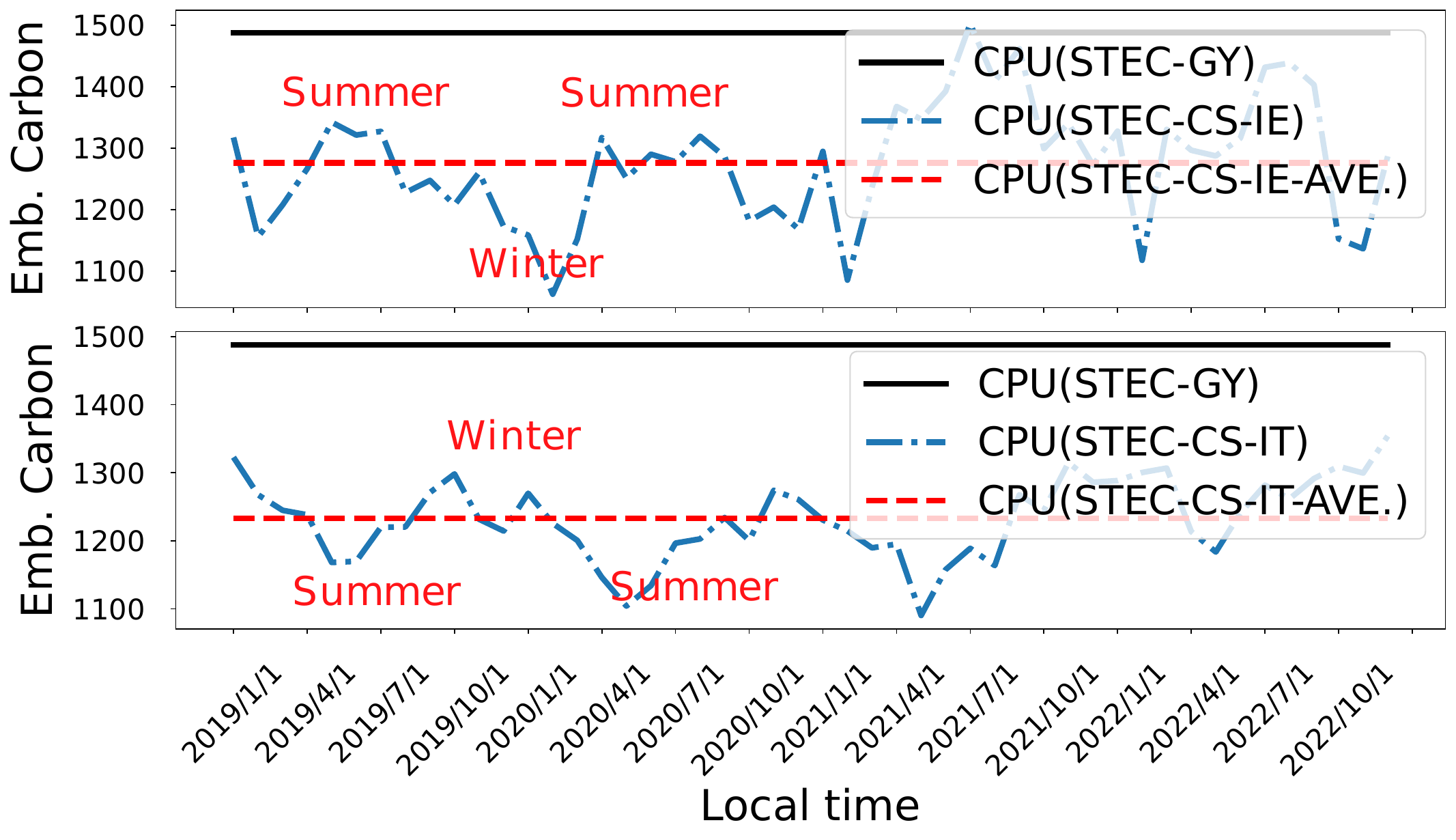}
    \vspace{-1.5em}
    \caption{The comparison between STEC-CS and STEC-GY. 
    }
    \label{fig: STEC-CS-timeline}
    \vspace{-1.5em}
\end{figure}

Table \ref{table: stec-cs} compares STEC-CS and STEC-GY (baseline) in the six major IC manufacturing regions. Fig. \ref{fig: STEC-CS-timeline} visually shows the embodied carbon ($g/cm^2$) output by the STEC-CD and STEC-GY in the CPU in Ireland and Italy from 2019 to 2022. We can find the following. (1) The difference is also significant (the average difference: 13.33\%, the average maximum difference: 27.50\%). For each hardware, the average difference is  13.69\% (CPU), 16.51\% (SSD), 10.36\% (HDD), and 12.74\% (Memory). (2) The embodied carbon has seasonal patterns in some regions. As Fig. \ref{fig: STEC-CS-timeline} shows, the embodied carbon is higher in summer and low in winter in Ireland, while, it shows the opposite trend in Italy. Intrinsically, the pattern depends on the climate and the composition of energy sources in the grids of a region. Specifically, wind power dominates in the variable renewable energy in Ireland, while, solar power dominates in the variable renewable energy in Italy. Besides, the wind source is abundant in the winter in Ireland, and the solar source is abundant in the summer in Italy. These together lead to different patterns in the two countries.

\begin{table}[]
\caption{The comparison between STEC-CS and STEC-GY on CPU (7nm), memorgy (10nm DDR4), SSD \cite{ssd}, and HDD \cite{hdd}.}
\vspace{-1.5em}
\label{table: stec-cs}
\begin{tabular}{ccc}
\toprule
Hardware & Ave. Difference (\%) & Max. Difference (\%) \\ \toprule
CPU & 13.69 & 27.32 \\
SSD & 16.51 & 34.46 \\
HDD & 10.36 & 21.63 \\
Memory & 12.74 & 26.60 \\ \hline
Average & 13.33 & 27.50 \\ \toprule
\end{tabular}
\vspace{-1.5em}
\end{table}

\subsection{The Evaluation of STEC-ZY}
Table \ref{table: stec-zy} presents the comparison between STEC-ZY and STEC-GY (baseline) in the two treaty zones (ASEAN and EU). Overall, the average difference is 18.00\%, and the average maximum difference is 19.29\%. We can find that (1) the STEC-CD/CS/ZY models all have significant differences with the STEC-GY (baseline). (2) Finer granularity can bring more dynamics. For example, the average maximum difference between STEC-CD and STEC-GY is greater than that between STEC-CS and STEC-GY, STEC-ZY and STEC-GY (33.62\% compared to 27.50\%, 33.62\% compared to 19.29\%).

\begin{table}[]
\caption{The comparison between STEC-ZY and STEC-GY on CPU (7nm), memory (10nm DDR4), SSD \cite{ssd}, and HDD \cite{hdd}.}
\vspace{-1.5em}
\label{table: stec-zy}
\scalebox{0.75}{
\begin{tabular}{cccccc}
\toprule
\multirow{2}{*}{Hardware} & \multicolumn{2}{c}{STEC-ZY} & STEC-GY & \multirow{2}{*}{\begin{tabular}[c]{@{}c@{}}Ave.\\ difference\\ (\%)\end{tabular}} & \multirow{2}{*}{\begin{tabular}[c]{@{}c@{}}Max.\\ difference\\ (\%)\end{tabular}} \\ \cline{2-4}
 & \begin{tabular}[c]{@{}c@{}}Emb.carbon\\ in EU\end{tabular} & \begin{tabular}[c]{@{}c@{}}Emb.carbon\\ in ASEAN\end{tabular} & Emb.carbon &  &  \\ \toprule
CPU & 1266.88 & 1848.10 & 1557.49 & 18.65 & 19.99 \\
SSD & 4.77 & 7.41 & 6.09 & 21.64 & 23.19 \\
HDD & 4.36 & 5.81 & 5.08 & 14.28 & 15.3 \\
Memory & 70.72 & 49.72 & 60.22 & 17.42 & 18.67 \\ \hline
Average & / & / & / & 18.00 & 19.29 \\ \toprule
\end{tabular}}
\vspace{-1.5em}
\end{table}

\subsection{A Case on Storm Malik Effect}

\begin{figure}[htbp]
    \centering
    \includegraphics[width=0.46\textwidth]{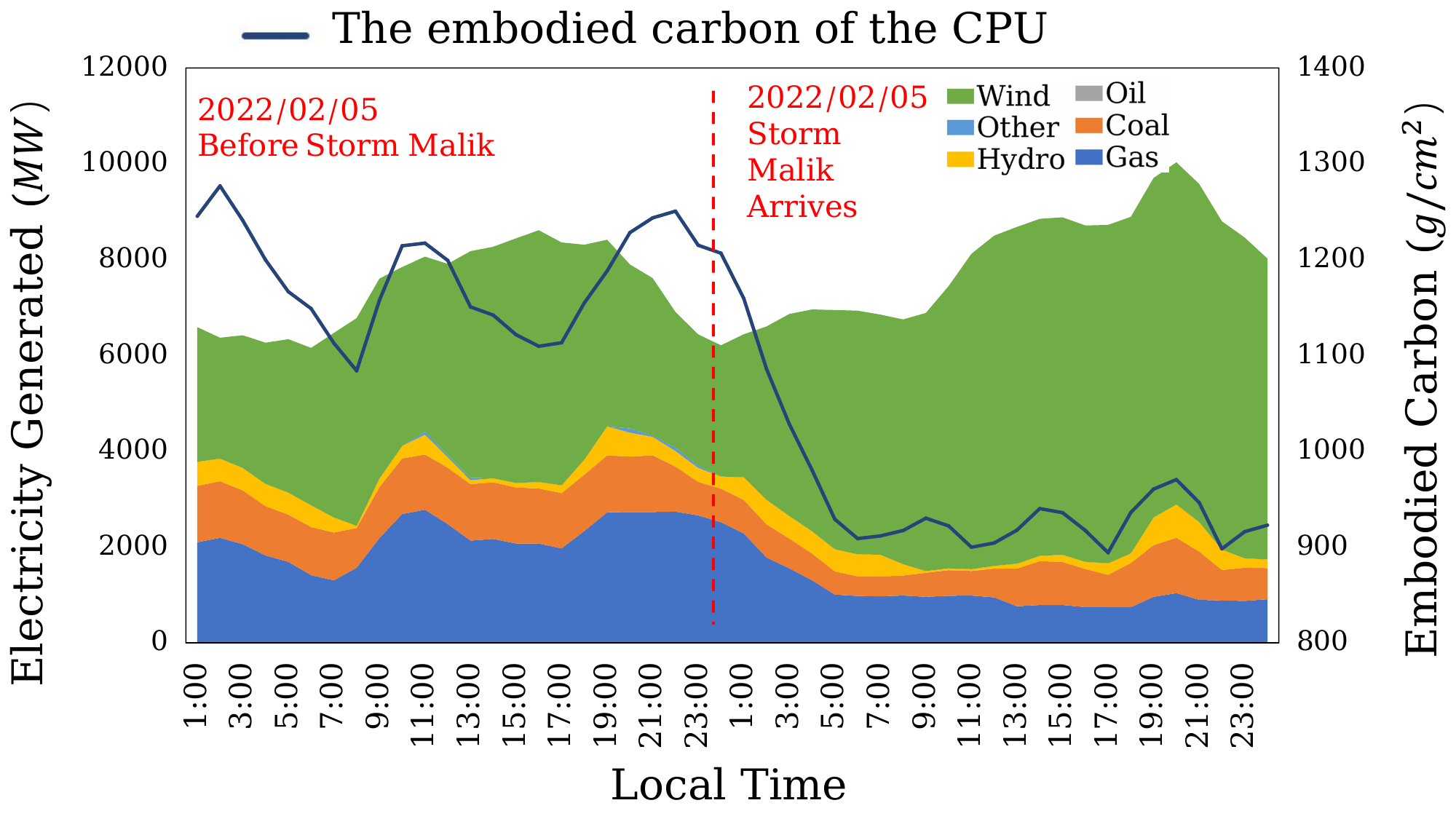}
    \vspace{-1.5em}
    \caption{
    Hourly embodied carbon of CPU and electricity generated in Ireland during Storm Malik.
    }
    \label{fig: storm}
    \vspace{-1.5em}
\end{figure}

Fig. \ref{fig: storm} visually presents dynamics in the embodied carbon of the CPU and the energy sources in grids in Ireland during the period of Storm Malik. For example, the embodied carbon reaches the maximum value (1276.87 $g/cm^2$) at 2 a.m. on the day before Storm Malik arrives. When Storm Malik arrives, the embodied carbon decreases rapidly, reaching a minimum (894.01 $g/cm^2$) at 5 p.m. The reason for this decline is that wind power increases rapidly under the influence of the storm. We can find that the embodied carbon has more variance in hourly time scale affected by weather. 

%% file: 6-Conclusion.tex
\section{Limitation and Future Work}

The key observation of this paper is that the carbon emission per unit of electricity generated at different locations and different time differs. As such, for the same product class, e.g., 28nm CPU, the carbon emissions of its product instances (e.g., made in the summer of Taiwan (TSMC) or in the winter of Ireland (Intel)) can differ. We present new spatial-temporal embodied carbon (STEC) models on key computer hardware, CPU, memory, and storage by extending a state-of-the-art embodied carbon model in the spatial-temporal dimension, and our evaluation validated significant differences.

There are several future works that can also address some of the limitations of this study: (1) in this paper, we omitted the cascading effect for the sake of simplicity. More specifically, the embodied carbon of a product is composed of the embodied carbon of upstream products. Upstream products can have their own spatial-temporal dynamics; and this leads to a cascading effect; (2) carbon emission data are important to develop embodied carbon models. Clearly, if a company can directly release the embodied carbon of its products in spatial-temporal dimensions, there is no need to develop STEC models. Unfortunately, data are scattered in various reports. In this paper, we made efforts to collect data, and our data collection was ad hoc and best-effort. It would be an important problem to develop a data report convention, i.e., what data could be expected and from whom. This asks for collective efforts; and (3) to investigate how spatial-temporal embodied carbon models in various granularity can assist real-world applications in embodied carbon reduction.

%% file: 7-Appendix.tex
\begin{appendix}

\section{Appendix}
To enable further investigation into the spatial-temporal of embodied carbon, we provide a supplementary for evaluation, detailed data sources for the STEC models, and describe the configurable parameters within the proposed STEC models.

\subsection{Supplementary for Evaluation}

\begin{figure}[htbp]
    \centering
    \includegraphics[width=0.46\textwidth]{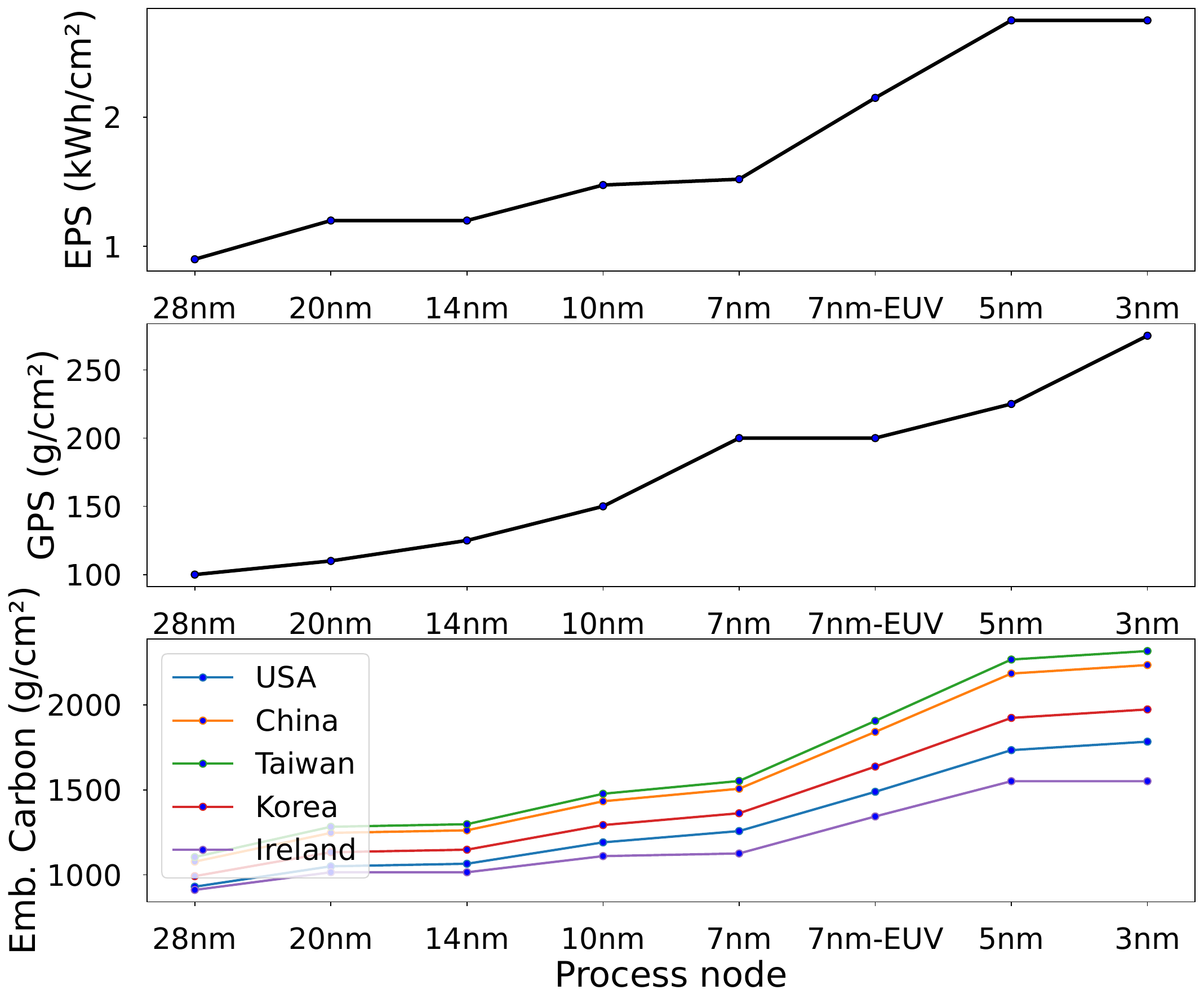}
    \vspace{-0.6em}
    \caption{
    Embodied carbon of CPUs varying from IC manufacturing technology nodes and production locations.
    }
    \label{fig: spitical}
    \vspace{-1em}
\end{figure}

 Fig. \ref{fig: spitical} presents the embodied carbon of CPU across technology nodes from 28nm to 3nm, considering different spatial information. The top and middle figures show the electricity consumed per unit size (EPS) and carbon released by the gas in the manufacturing process rise but do not vary according to space, as illustrated before. The bottom figure presents the embodied carbon across technology nodes in five major IC-producing regions. We can clearly see that more advanced technology nodes will further increase the spatial disparity of embodied carbon.

\begin{table}[htbp]
\caption{Carbon intensity of electricity in six major IC production regions in 2022.}
\label{table: CI_2022}
\begin{tabular}{cc}
\toprule
Countriy/Region & \begin{tabular}[c]{@{}c@{}}Carbon intensity of electricity\\ (g/kWh)\\ (\textcolor{blue}{Date Sources: Our World in Data \cite{WorldinData}})\end{tabular} \\ \toprule
Taiwan & 561 \\
China & 531 \\
South Korea & 436 \\
United States & 367 \\
Italy & 372 \\
Ireland & 346 \\ \toprule
\end{tabular}
\end{table}

\begin{table}[htbp]
\caption{Carbon emission factors (g/kWh) for different energy sources}
\scalebox{1}{
\label{table: ef}
\begin{tabular}{lc}
\toprule
\multicolumn{1}{c}{Energy sources} & \begin{tabular}[c]{@{}c@{}}Direct Emission Factors\\ (\textcolor{blue}{Data Source: Research Paper \cite{maji2022carboncast}})\end{tabular} \\ \toprule
Oil                                & 406                                                                                            \\
Coal                               & 760                                                                                            \\
Natural gas                        & 370                                                                                            \\
Nuclear                            & 0                                                                                              \\
Wind                               & 0                                                                                              \\
Solar                              & 0                                                                                              \\
Hydro                              & 0                                                                                              \\
Geothermal                         & 0                                                                                              \\
Biomass                            & 0                                                                                              \\
Other                              & 575                                                                                            \\ \toprule
\end{tabular}}
\end{table}

\begin{table}[]
\caption{Embodied carbon parameters in semiconductor manufacturing}
\scalebox{0.75}{
\label{table: eps_processor}
\begin{tabular}{cccc}
\toprule
    Process Node & \begin{tabular}[c]{@{}c@{}}EPS (g/cm2)\\ (\textcolor{blue}{Data Source:} \\ \textcolor{blue}{Research Paper \cite{bardon2020dtco}})\end{tabular} & \begin{tabular}[c]{@{}c@{}}GPS (g/cm2)\\ (\textcolor{blue}{Data Source:} \\ \textcolor{blue}{Research Paper \cite{bardon2020dtco}})\end{tabular} & \begin{tabular}[c]{@{}c@{}}MPS (g/cm2)\\ (\textcolor{blue}{Data Source:} \\ \textcolor{blue}{Environmental Report \cite{boyd2011life}})\end{tabular} \\ \toprule
28 & 0.9 & 100 & 500 \\
20 & 1.2 & 110 & 500 \\
14 & 1.2 & 125 & 500 \\
10 & 1.475 & 150 & 500 \\
7 & 1.52 & 200 & 500 \\
7-EUV & 2.15 & 200 & 500 \\
7-EUV-DP & 2.15 & 200 & 500 \\
5nm & 2.75 & 225 & 500 \\
3nm & 2.75 & 275 & 500 \\ \toprule
\end{tabular}}
\end{table}

\begin{table}[]
\caption{Embodied carbon, bit density, and carbon emissions from electricity consumption of memory \cite{Jeongdong2017,gupta2022act}.}
\scalebox{0.8}{
\begin{tabular}{lccc}
\toprule
\multicolumn{1}{c}{Technology} & \begin{tabular}[c]{@{}c@{}}Embodied Carbon\\ (g/GB)\end{tabular} & \begin{tabular}[c]{@{}c@{}}Bit Density\\ (G/mm2)\end{tabular} & \begin{tabular}[c]{@{}c@{}}Carbon Emissions from \\ Electricity Consumption\\ (g/GB)\end{tabular} \\ \toprule
30nm LPDDR3 & 230 & 0.06 & 67.50 \\
20nm LPDDR3 & 184 & 0.11 & 51.43 \\
10nm DDR4 & 65 & 0.19 & 35.74 \\
LPDDR4 & 48 & 0.17 & 39.04 \\ \toprule
\end{tabular}}
\label{table: appmem}
\end{table}

\subsection{Data sources for STEC Models}
Table \ref{table: datasource} presents the data sources for STEC models. All the data are public. Table \ref{table: datasource} (a) shows the sources for the electricity data, including the carbon intensity (yearly and monthly) and energy sources (hourly). Table \ref{table: datasource} (b) shows the sources for the hardware-related parameters.

\begin{table*}[htbp]
\caption{The Data Sources for the STEC models}
\label{table: datasource}
\scalebox{0.8}{
\begin{tabular}{clcccccccl}
\toprule
\multicolumn{10}{c}{(a) Data Sources for Electricity Energy} \\ \toprule
 & \multicolumn{1}{c}{EU} & ASEAN & China & Taiwan & South Korea & USA & Ireland & Italy & \multicolumn{1}{c}{Source} \\ \hline
Yearly {[}2019-2022{]} & \multicolumn{1}{c}{\checkmark} & \checkmark & \checkmark & \checkmark & \checkmark & \checkmark & \checkmark & \checkmark & \textcolor{blue}{Our Word in Data \cite{WorldinData}}\\
Monthly {[}2019-2022{]} & \multicolumn{1}{c}{-} & - & \checkmark & \checkmark & \checkmark & \checkmark & \checkmark & \checkmark & \textcolor{blue}{EMBIR \cite{embir}} \\
Daily {[}2021{]} & \multicolumn{1}{c}{-} & - & - & - & - & - & \checkmark & \checkmark & \textcolor{blue}{ENTSOE \cite{ENTSOE}} \\ \toprule
\multicolumn{10}{c}{(b) Data Sources for Other Types of Carbon Emission} \\  \toprule
Parameter & \multicolumn{6}{c}{Description} & \multicolumn{2}{c}{Unit} & \multicolumn{1}{c}{Source} \\ \hline 
EPS & \multicolumn{6}{l}{Electricity consumed per die Size} & \multicolumn{2}{c}{kWh/cm2} & \textcolor{blue}{Research paper \cite{bardon2020dtco}} \\
GPS & \multicolumn{6}{l}{Carbon emission from Gas used per die Size} & \multicolumn{2}{c}{g/cm2} & \textcolor{blue}{Research paper \cite{bardon2020dtco}} \\
MPS & \multicolumn{6}{l}{Carbon emission from Material used per die Size} & \multicolumn{2}{c}{g/cm2} & \textcolor{blue}{Industrial research reports \cite{boyd2011life}} \\
BD & \multicolumn{6}{l}{Bit density for memory} & \multicolumn{2}{c}{GB/cm2} & \textcolor{blue}{Industrial research reports \cite{Jeongdong2017}} \\ 
EPG & \multicolumn{6}{l}{Electricity consumed per GB} & \multicolumn{2}{c}{kWh/GB} & \begin{tabular}[c]{@{}l@{}} \textcolor{blue}{Industrial environmental}\\ \textcolor{blue}{reports of manufacturers \cite{SK,storage}}  \end{tabular} \\ \toprule
\end{tabular}}
\end{table*}

\subsection{Embodied Carbon Parameters}
Table \ref{table: CI_2022} presents the annual average carbon intensity of electricity in six major IC production regions in 2022. Table \ref{table: ef} summarizes the direct carbon emissions factors for each energy source. Table \ref{table: eps_processor},  \ref{table: appmem}, and \ref{table: storage} summarize the related embodied carbon data for CPU, memory, and storage, respectively.

\begin{table*}[]
\caption{The summary of the embodied carbon of storage from product reports}
\label{table: storage}
\scalebox{0.7}{
\begin{tabular}{|c|c|c|c|c|c|}
\hline
Types                           & Technology                   & Embodied Carbon (g/GB) & Manufacturing Energy Carbon (g/GB) & Other Carbon (g/GB) & Date Source \\ \hline
\multirow{4}{*}{Enterprise SSD} & Nytro 3530                   & 6.27                   & 4.25                               & 2.02                & \textcolor{blue}{LCA Report \cite{ssd}}           \\ \cline{2-6} 
                                & Nytro 1551                   & 3.91                   & 1.53                               & 2.38                & \textcolor{blue}{LCA Report \cite{b}}           \\ \cline{2-6} 
                                & Nytro 3331                   & 5.48                   & 0.92                               & 4.56                & \textcolor{blue}{LCA Report \cite{c}}           \\ \cline{2-6} 
                                & Nytro 3332                   & 2.42                   & 0.78                               & 1.64                & \textcolor{blue}{LCA Report \cite{d}}           \\ \hline
Consumer SSD                    & BarraCuda 120 SSD            & 26.28                  & 23.85                              & 2.43                & \textcolor{blue}{LCA Report \cite{f}}           \\ \hline
\multirow{9}{*}{Enterprise HDD} & EXOS X20                     & 0.88                   & 0.36                               & 0.52                & \textcolor{blue}{LCA Report \cite{g}}           \\ \cline{2-6} 
                                & EXOS X18                     & 0.88                   & 0.39                               & 0.49                & \textcolor{blue}{LCA Report \cite{h}}           \\ \cline{2-6} 
                                & Exos 2X14                    & 1.28                   & 0.51                               & 0.78                & \textcolor{blue}{LCA Report \cite{i}}           \\ \cline{2-6} 
                                & Exos 7E8                     & 5.28                   & 2.34                               & 2.94                & \textcolor{blue}{LCA Report \cite{hdd}}           \\ \cline{2-6} 
                                & Exos 5E8                     & 2.54                   & 1.14                               & 1.40                & \textcolor{blue}{LCA Report \cite{k}}           \\ \cline{2-6} 
                                & Exos 10E2400                 & 10.75                  & 6.94                               & 3.81                & \textcolor{blue}{LCA Report \cite{l}}           \\ \cline{2-6} 
                                & EXOS 15E900                  & 21.62                  & 10.65                              & 10.97               & \textcolor{blue}{LCA Report \cite{m}}           \\ \cline{2-6} 
                                & Exos X16                     & 1.46                   & 0.77                               & 0.69                & \textcolor{blue}{LCA Report \cite{n}}           \\ \cline{2-6} 
                                & Exos X12                     & 1.32                   & 0.53                               & 0.79                & \textcolor{blue}{LCA Report \cite{o}}           \\ \hline
\multirow{11}{*}{Consumer HDD}  & BarraCuda 3.5                & 9.40                   & 4.84                               & 4.56                & \textcolor{blue}{LCA Report \cite{p}}           \\ \cline{2-6} 
                                & BarraCuda                    & 4.25                   & 2.08                               & 2.17                & \textcolor{blue}{LCA Report \cite{q}}           \\ \cline{2-6} 
                                & BarraCuda Pro                & 2.62                   & 1.22                               & 1.40                & \textcolor{blue}{LCA Report \cite{r}}           \\ \cline{2-6} 
                                & FireCuda                     & 5.16                   & 3.81                               & 1.35                & \textcolor{blue}{LCA Report \cite{s}}           \\ \cline{2-6} 
                                & IronWolf                     & 5.28                   & 2.22                               & 3.06                & \textcolor{blue}{LCA Report \cite{t}}           \\ \cline{2-6} 
                                & IronWolf Pro                 & 3.80                   & 1.33                               & 2.47                & \textcolor{blue}{LCA Report \cite{u}}           \\ \cline{2-6} 
                                & Skyhawk 3 TB                 & 9.85                   & 2.17                               & 7.68                & \textcolor{blue}{LCA Report \cite{v}}           \\ \cline{2-6} 
                                & Skyhawk Surveillance HDD     & 4.37                   & 1.54                               & 2.83                & \textcolor{blue}{LCAReport \cite{w}}           \\ \cline{2-6} 
                                & Skyhawk 6 TB                 & 4.18                   & 1.09                               & 3.09                & \textcolor{blue}{LCA Report \cite{x}}           \\ \cline{2-6} 
                                & Video 3.5 HDD                & 8.20                   & 3.22                               & 4.98                & \textcolor{blue}{LCA Report \cite{y}}           \\ \cline{2-6} 
                                & Video 3.5 HDD (Pipeline HDD) & 9.54                   & 3.23                               & 6.31                & \textcolor{blue}{LCA Report \cite{z}}           \\ \hline
\multirow{2}{*}{External HDD}                    & ULTRA TOUCH                  & 5.54                   & 3.40                               & 2.13                & \textcolor{blue}{LCA Report \cite{qq}}          \\ \cline{2-6} 
                                & Rugged Mini                  & 4.22                   & 2.98                               & 1.25                & \textcolor{blue}{LCA Report \cite{xx}}          \\ \hline
\end{tabular}}
\end{table*}

\end{appendix}